\documentclass[pre,aps,twocolumn,superscriptaddress]{revtex4-1}

\usepackage{graphicx}
\usepackage{bm}
\usepackage{amsfonts}
\usepackage{color}
\usepackage{hyperref}
\usepackage{epstopdf, epsfig}
\usepackage{tikz}

\newcommand{\de}{\mathrm{d}}
\newcommand{\xp}{\bm x_\mathrm{p}}
\newcommand{\vp}{\bm v_\mathrm{p}}
\newcommand{\taup}{\tau_\mathrm{p}}
\newcommand{\U}{U}
\newcommand{\ds}{d}
\newcommand{\Rey}{\textit{Re}}
\newcommand{\St}{\textit{St}}
\newcommand{\Fr}{\textit{Fr}}
\newcommand{\Sg}{\textit{Sv}}
\newcommand{\dd}[2]{\frac{\de{#1}}{\de{#2}}}
\newcommand{\mycirc}[1]{%
   \begin{tikzpicture}[baseline=(C.base)]
     \node[draw,circle,inner sep=0.5pt](C) {#1};$\!$
   \end{tikzpicture}}

\newcommand{\cemefaddress}{MINES ParisTech, PSL Research University,
  CNRS, CEMEF, Sophia-Antipolis, France}
\newcommand{\inriaaddress}{Universit\'e C\^ote d'Azur, INRIA, CNRS, CEMEF, Sophia-Antipolis, France}

\graphicspath{{./Figs/}}

\begin{document}

\title{Effect of gravitational settling on the collisions of small inertial particles
  with a sphere}

\author{J\'er\'emie Bec} \affiliation{\inriaaddress}\affiliation{\cemefaddress}
\author{Christoph Siewert} \affiliation{Deutscher Wetterdienst (DWD),
  Offenbach, Germany}
\author{Robin Vall\'{e}e} \affiliation{\cemefaddress}
\begin{abstract}
  The rate at which small inertial particles collide with a
  moderate-Reynolds-number spherical body is found to be strongly
  affected when the formers are also settling under the effect of
  gravity. The sedimentation of small particles indeed changes the
  critical Stokes number above which collisions occur. This is
  explained by the presence of a shielding effect caused by the
  unstable manifolds of a stagnation-saddle point of an effective
  velocity field perceived by the small particles.  It is also found
  that there exists a secondary critical Stokes number above which no
  collisions occur. This is due to the fact that large-Stokes number
  particles settle faster, making it more difficult for the larger one to
  catch them up.  Still, in this regime, the flow disturbances create
  a complicated particle distribution in the wake of the collector,
  sometimes allowing for collisions from the back.  This can lead
  to collision efficiencies higher than unity at large values of the
  Froude number.
\end{abstract}

\maketitle

\section{Introduction}
The capture of small suspended particles by a streamlined or bluff
body is an important process in many natural systems.  Besides wind
pollinisation~\cite{niklas1987pollen} or the collection of
phytoplancton by passive suspension-feeding invertebrates
\cite{riisgaard2010particle}, it plays a crucial role in planet
formation~\cite{lissauer1993planet}, in particular when estimating the
growth rate of planetesimals by sweep-up and accretion of small-size
dust grains~\cite{windmark2012planetesimal}. Several atmospheric
processes also involve the capture of small particles by a larger drop
or ice crystal, and estimating such phenomena is key in the
parameterization of cloud-resolving meteorological
models~\cite{pruppacher1997microphysics}.  Collection rates are for
instance needed when accounting for the growth of raindrops by
accretion of cloud droplets~\cite{seifert2001double}, for the riming
of supercooled droplets by ice crystals~\cite{wang2000collision}, and
for the scavenging of aerosols during wet
deposition~\cite{mircea2000precipitation}.  Capturing particles in
dirty gases is also an important industrial challenge. Most techniques
rely on the collection of particles by water drops, such as in reverse
jet scrubbers~\cite{lim2006prediction}.  In all these applications,
achieving precise estimates requires, on the one hand, elucidating
mesoscopic fluid-dynamical effects that determine whether or not
impaction occurs, and on the other hand, specifying the microphysical
features and processes that affect the outcome of such collisions and
a possible capture by the collector~\cite{spielman1977particle}.

A large object moving across a fluid creates a flow that pushes fluid
elements away from its surface.  Small-size particles uniformly
suspended in the fluid are then drifted aside, so that the rate at
which they are collected is less than the ideal rate obtained by
considering the volume swept by the large object.  The fraction of
particles that actually collide defines the \textit{collision
  efficiency}, a quantity that enters most kinetic models.  An
efficiency larger than zero requires that particles detach from the
fluid streamlines. This can be brought about by several effects,
including Brownian motion, boundary interception, and
inertia~\cite{carotenuto2010wet}.  While the formers can be treated
analytically by probabilistic and geometric arguments, inertial
impaction is still mainly addressed with empirical approaches.  By
fitting numerical and experimental measurements at moderate values of
the large-object Reynolds number, one obtains formulae expressing the
collision efficiency as a function of the small-particle stopping time
~\cite{beard1974numerical,slinn1974precipitation} that are today at
the basis of model parameterization.

Much work has been recently devoted to improving collision
efficiencies and providing refined statistics that are of importance
to the collisional processes. Many aspects have been covered,
including the flow modifications due to the collected
particles~\cite{wang2005theoretical,chen2018turbulence}, the
statistics of impact velocities~\cite{mitra2013can}, fluctuations in
the particle concentration~\cite{homann2015concentrations}, the effect
of a large Reynolds number of the collector~\cite{wang2016study}, the
presence of turbulent fluctuations in the surrounding
fluid~\cite{homann2016effect,aarnes2019inertial}, the outcomes of
elastic rebounds~\cite{vallee2018inelastic}, and the fluid-structure
interaction between the collector and the
flow~\cite{mccombe2018collector}.  However an effect of particular
importance to atmospheric applications has been neglected so
far. Usually, while the collecting object (raindrop or ice crystal)
falls through the fluid under the effect of gravity, the collected,
small-size particles are themselves settling and decouple from the
fluid. Gravitational settling is known to have drastic impacts on the
dynamics and collisions between small inertial particles, in
particular when the carrier flow is
turbulent~\cite{ayala2008effects,bec2014gravity}. Except in specific
cases related to aerosol
washout~\cite{beard1974numerical,pitter1974numerical,pruppacher1997microphysics},
not much is known on the way settling affects collision efficiencies
due to inertial impaction.

We consider here the fundamental problem of a large-size spherical
object freely falling in an incompressible flow at rest with a small
or moderate Reynolds number. On its way it collects small-size heavy
inertial particles that themselves settle at lower terminal
speeds. Using numerical simulations and phenomenological arguments we
quantify the collision efficiency as a function of the three
dimensionless parameters characterizing the problem, namely the
large-particle Reynolds number $\Rey$, the small particle Stokes
number $\St$, and the Froude number $\Fr$, that measures the
importance of the involved hydrodynamical forces with respect to
gravity.  As already known in the case without gravity, inertial
impaction occurs only if the small-particle Stokes number is large
enough. We find that gravity leads to a shielding effect that
increases the corresponding critical Stokes number. Furthermore, in
the presence of gravity, as already observed
in~\cite{pitter1974numerical}, there exists a second critical Stokes
number above which small particles fall so fast that they are never
collected by the sphere.  Concretely, this means that inertial
impaction can only occur when the small-particle sizes belong to a
specific window. We moreover find that in this window, efficiencies
can be larger than unity, meaning that the sphere actually collect
more particles than those present in the volume it sweeps. This
surprising phenomenon can be explained by the accumulation of small
particles in the wake of the large sphere where they are
entrained. Thanks to gravity, they can then catch up with the sphere
and collide on its tail.

The paper is organized as follows. In Sec.~\ref{sec:model} we
introduce the model, the important parameters and observables, and our
methodology. Section~\ref{sec:efficiency} is dedicated to the behavior
of the collision efficiency in the specific case where the large
particle has a negligible Reynolds number. In
Section~\ref{sec:shielding}, we present phenomenological arguments
that explain the shielding effect and the observed behavior of
the critical Stokes numbers. Section~\ref{sec:caustics} reports
observations and interpretations on backward collisions and their link
with the creation of caustics and concentrations in the wake of the
sphere. In Sec.~\ref{sec:reynolds} we extend such considerations to
the case of a finite Reynolds number of the large sphere and find that
our results stay there valid. Finally, section~\ref{sec:conclusion}
encompasses concluding remarks and perspectives.

\section{Model and parameters}
\label{sec:model}

We consider a large spherical particle with diameter $\ds$ immersed in
a three-dimensional incompressible fluid whose dynamics solves the
Navier--Stokes equation.  The flow is assumed at rest at infinity and
the fluid velocity field obeys a no-slip boundary condition at the
surface of the sphere.  The spherical particle represents a
collector. It moves with a steady speed $U$ obtained by balancing
gravity, buoyancy and the drag exerted by the fluid.  Without loss of
generality, we work in the reference frame attached to the sphere and
whose origin is at its center. The fluid flow is thus at rest at the
particle surface, namely the velocity field satisfies
$\bm u(\bm x,t) = 0$ at $|\bm x|=\ds/2$. It tends to $U\bm e_z$ when
$|\bm x| \to\infty$.

Small heavy particles are suspended in the fluid.  Their positions
$\xp$ and velocities $\vp$ follow the dynamics
\begin{eqnarray}
  &&\dd{\xp}{t} = \vp, \nonumber \\
  &&\dd{\vp}{t} = \beta \frac{\mathrm{D}\bm u}{\mathrm{D}t}(\xp,t)
     -\frac{1}{\taup}\!\left[\vp - \bm u(\xp,t)
     \right]+(1\!-\!\beta)\,\bm g,
     \label{eq:dynpart}
\end{eqnarray}
where
$\beta =
3\,\rho_{\mathrm{f}}/(2\,\rho_{\mathrm{p}}+\rho_{\mathrm{f}})$ is the
added mass factor, $\taup = a^2 /(3\beta\nu)$ the particle response
time, $a$ designating its radius, $\rho_{\mathrm{p}}$ its mass
density, $\rho_{\mathrm{f}}$ and $\nu$ the fluid density and kinematic
viscosity, respectively, and $\bm g = -g\,\bm e_z$ is the acceleration
of gravity.  The particle sizes are assumed sufficiently small to
consider only three forces in the right-hand side of
(\ref{eq:dynpart}), namely the added-mass, the viscous drag, and
buoyancy, and to neglect the Basset-Boussinesq history term and
Fax\'en's finite-size corrections. In addition, particles are
sufficiently dilute to neglect their possible feedback onto the fluid.
We moreover suppose that they are uniformly distributed at
$z\to -\infty$ with a velocity equal to their terminal speed, namely
$\vp = (1\!-\!\beta)\taup\,\bm g +\U\,\bm e_z$ in the reference frame
of the large spherical particles.

One usually writes the system in a non-dimensional form by expressing
length scales in units of the particle diameter $d$ and timescales in
terms of the sweeping time $d/U$. This leads to
introduce three non-dimensional parameters:\\
-~the \textit{Reynolds number} $\Rey = \U\,\ds/\nu$, which characterizes
the flow around the large sphere,\\
-~the \textit{Froude number} $\Fr = \U^2/(\ds\,g)$, which measures the
importance of gravity,\\
-~the \textit{Stokes number} $\St = \taup\,\U/\ds$, which quantifies the
small-particles inertia.

The Stokes number, together with the added mass factor $\beta$, are
specified by the nature of the collected particles.  We consider them
as variable parameters, because in most applications, such particles
are polydisperse and have a broad distribution of sizes and masses.
The Stokes number measures the inertia of the small particles. When
$St=0$, they behave as tracers, exactly follow the fluid streamlines,
and collide with the large object only by diffusion or
interception. When $St\to\infty$, the particles completely detach from
the flow and collide with the collector from the moment that they are
located on its path. In the presence of gravity, the small particles
are themselves settling. The importance of gravity with respect to
inertia is often measured in terms of the gravitational Stokes number
$\Sg = (1\!-\!\beta)\taup\,g/\U = (1\!-\!\beta)\St/\Fr$ (see, e.g.,
\cite{ayala2008effects}), which corresponds in our case to the ratio
between the terminal velocities of the small particles and that of the
large sphere.  Clearly, when $\Sg>1$, the inertial particles settle
faster than the sphere and collisions occur in the wake.

\begin{figure}[t]
  \begin{center}
    \includegraphics[width=\columnwidth]{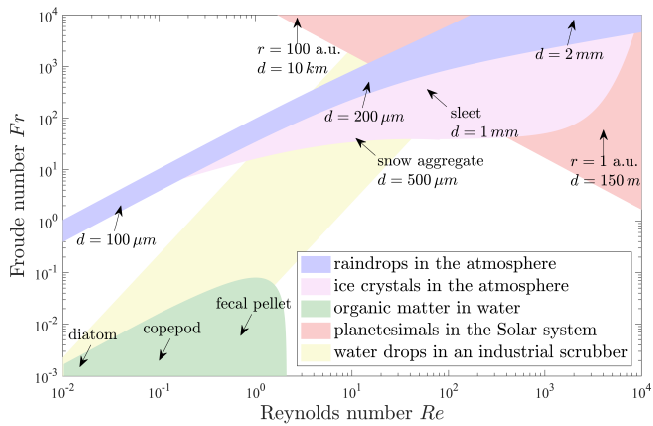}
  \end{center}
  \vspace{-18pt}
  \caption{\label{fig:sketch_appli} Typical ranges of Reynolds and
    Froude numbers encountered in applications. Different examples
    illustrate typical sphere radii and settings.}
\end{figure}

The Reynolds and the Froude numbers depend upon the considered
application. Figure~\ref{fig:sketch_appli} sketches the typical ranges
covered by these parameters in atmospheric physics, in oceanology, in
planet formation and in industrial scrubbers. In the case of raindrops
falling in the atmosphere, the velocity $U$ is given by the large drop
terminal speed, $g$ is the near-Earth-surface acceleration of gravity
and the observed spreading comes from variations in pressure,
temperature, and drop's shape, which tends to become more oblate when
its diameter increases above $1\,mm$~(see, e.g.,
\cite{feng2011raindrop}).  Settling crystals define a broader range of
parameters depending on meteorological conditions and how densely ice
is packed.  Organic matter in the ocean, such as phytoplankton, has a
mass density very close to water and is thus settling at a rather low
speed~\cite{lee2009particulate}, whence small values of the parameters
$\Rey$ and $\Fr$. Understanding the rate at which the larger particles
accrete smaller sediments is important to quantify the downward
piggy-back transport~\cite{lal1980comments}, and thus the efficiency
of the CO$_2$ oceanic pump~\cite{alonso2010role}.  For planet
formation, gravity depends upon the distance $r$ between the
planetesimals and the star, which is expressed here in astronomical
units (a.u.). The protoplanetary disk is made of gas whose
thermodynamical properties depend on $r$. The radial pressure of the
disk maintains the gas at a sub-Keplerian orbiting speed. The
planetesimals, whose sizes range from several meters to hundreds of
kilometers, have a drag with the gas and slowly drift inward at
velocities of the order of tens of meters per
seconds~\cite{homann2016effect}. On their path, they collect
additional dust to eventually reach the sizes of planetary
embryos~\cite{guillot2014filtering}. Finally, in the case of
industrial wet scrubbers, a jet of water droplets with sizes of the
order of hundreds of microns is sent against a flow of polluted gas in
order to collect suspended particulate matter. The slip velocity $U$
of droplets depends upon the distance from the jet's nozzle.

In all these applications, we aim at understanding the efficiency with
which the large spherical collector accretes smaller particles.  In
the absence of fluid flow, the rate at which collisions occur is
obtained by considering the number of small particles contained in the
volume swept by the large sphere per unit time. This leads to the
``ideal'' collision rate
$Q_{\mathrm{no\ fluid}} = (\pi\,d^2/4)\,U\,n$, where $n$ designates
the small-particles number density that is assumed uniform.  In the
presence of a fluid flow, small particles are possibly drifted away
from the sphere, so that the actual collision rate
$Q_{\mathrm{fluid}}$ differs from the above estimate. Such a
discrepancy is measured in terms of the \textit{collision efficiency}
defined as $\mathcal{E} = Q_{\mathrm{fluid}}/Q_{\mathrm{no\
    fluid}}$. This quantity, which enters in all model
parametrizations, is our main observable. We aim at understanding
how it depends both on the features of the small accreted particles
($\beta$ and $\St$) and on the collector parameters ($\Rey$ and
$\Fr$).

We focus here on moderate values of the falling-sphere Reynolds number
$\Rey$, so that the perturbed flow is steady and
axisymmetric~\cite{fabre2008bifurcations}. This allows for a more
systematic investigation as a function of the other parameters.  We
first study the case when the flow has no inertia ($\Rey=0$). The
velocity field is given by the Stokes' equation, for which an explicit
analytical solution is known (see, e.g.,
\cite{falkovich2011fluid}). By comparing to the results of direct
numerical simulations, we find that, as expected, this simplified case
reproduces all the qualitative features observed for
$\Rey\lesssim 15$. Simulations are performed using the CimLib CFD
finite-element code~\cite{hachem2010stabilized}, which is able to
solve the Navier--Stokes equations with an arbitrary immersed object
using an adaptive meshing~\cite{billon2017anisotropic}.

\section{Collision efficiency}
\label{sec:efficiency}

\begin{figure}[b]
  \begin{center}
    \includegraphics[width=\columnwidth]{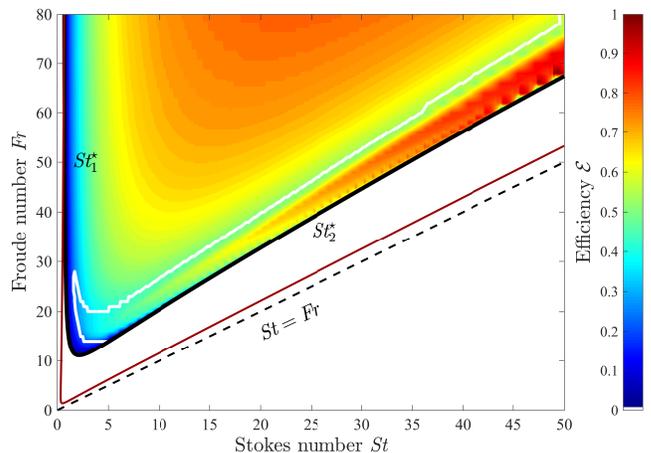}
  \end{center}
  \vspace{-18pt}
  \caption{\label{fig:efficiency_St_Fr} Collision efficiency as a
    function of the small-particles Stokes number $\St$ and of the
    Froude number $\Fr$ for the Stokes flow $\Rey=0$ and for
    $\beta=0$. The black dashed line $\Fr=\St$ corresponds to a
    settling Reynolds number of unity. The white curve in the colour
    area delimits the parameter region where backward collisions
    occur. The bold black curve is the fit (\ref{eq:fit_stcr}) to the
    critical line. The red curve corresponds to the bounds
    (\ref{eq:cond1})-(\ref{eq:cond2}) obtained in the next section.}
\end{figure}
In this section, we neglect the fluid inertia and fix the Reynolds
number to $\Rey=0$. This allows for a general overview on how the
collision efficiency $\mathcal{E}$ depends on the two parameters $\St$
and $\Fr$, as shown in Fig.~\ref{fig:efficiency_St_Fr} for very heavy
small-size particles ($\beta=0$). A first observation is that
$\mathcal{E}>0$ only for values of the Stokes and the Froude numbers
in a set with a triangular shape. In other words, collisions occur
only when $\Fr>\Fr^\star\approx 11$, and for a fixed Froude number
above this value, the efficiency is non-zero only for Stokes numbers
in the interval $\St_1^\star(\Fr) < \St < \St_2^\star(\Fr)$.  When
$\Fr\to\infty$, the lower critical Stokes number $\St_1^\star$
approaches from above the value
$\St^\star_{\mathrm{no\ grav}}\approx 0.605$ already identified in the
absence of
gravity~\cite{slinn1974precipitation,vallee2018inelastic}. This means
that gravitational settling tends to increase its value. The upper
critical Stokes number (above which no collisions occur) tends
asymptotically to infinity as $\St_2^\star \propto \Fr$ when
$\Fr\to\infty$. This critical Froude number is bounded from below by
the line $\Sg = (1\!-\!\beta)\St/\Fr = 1$. As stated above, when
$\Sg>1$, the small particles settles faster than the collector and
collide with it from the back.  Such settings could for instance be
encountered when interested in the collection of small particles by a
rising spherical bubble. Such settings are beyond the scope of this
work.

As can be seen from Fig.~\ref{fig:efficiency_St_Fr}, the critical
Stokes numbers $\St_1^\star$ and $\St_2^\star$ actually define a
single curve in the $(\St,\Fr)$ parameter space. An approximation is
more conveniently found in the $(\St,\Sg)$ space by looking for a fit
of the form
\begin{equation}
  \St^\star(\Sg) = \frac{\St^\star_{\mathrm{no\
        grav}}}{f(\beta_1\,\Sg)+\beta_2\,\Sg}, \label{eq:fit_stcr}
\end{equation}
where we use the function $f(x) = (1-7x^{2/3}/3-5x/3)/(1+3x/2)$, and
$\St^\star_{\mathrm{no\ grav}}$ is the critical Stokes number below
which no collisions occur in the absence of gravity. The constants
$\beta_1$, $\beta_2$ are two adjustable parameters that possibly
depend on the Reynold number. The black curve shown in
Fig.~\ref{fig:efficiency_St_Fr} was obtained by choosing
$\beta_1 = 1.38$ and $\beta_2=0.014$.  It gives a very good
approximation of the critical curve. As we will see in
Sec.~\ref{sec:reynolds}, this form can also be used to fit the
measurements made at finite values of the Reynolds number.

Another observation from Fig.~\ref{fig:efficiency_St_Fr} is the
presence of a second maximum of the collision efficiency close to the
boundary $\St=\St_2^\star$. This elongated region leads to
efficiencies higher than unity at large values of both $\Fr$ and
$\St$. Such a surprising behavior comes from collisions by small
particles from the back of the sphere. Such collisions are present in
the hook-shaped parameter-space region delimited by a white curve in
Fig.~\ref{fig:efficiency_St_Fr}.

\begin{figure}[ht]
  \begin{center}
    \includegraphics[width=\columnwidth]{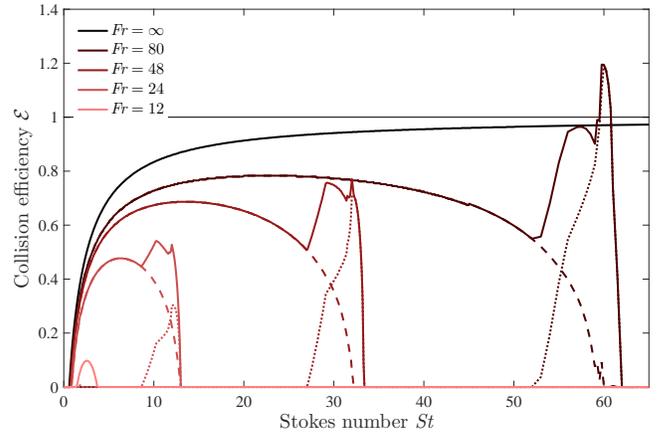}
  \end{center}
  \vspace{-18pt}
  \caption{\label{fig:efficiency_St} Collision efficiencies for
    $\Rey=0$, $\beta=0$, as a function of the small-particles Stokes
    number and different values of the Froude number as labelled.
    Solid lines correspond to the total efficiency obtained as the sum
    of the contribution from forward collisions (dashes) and
    backward collisions (dots).}
\end{figure}

Figure~\ref{fig:efficiency_St} represents four horizontal cuts of the
previous figure at different values of the Froude number.  We separate
there contributions from forward and backward collisions.  For an
infinite value of $\Fr$, that is in the absence of gravity, all
collisions occur on the head-on hemisphere of the large particle and
the obtained efficiency is that known for Stokes flow around a sphere,
which grows linearly from zero at
$\St = \St_{\mathrm{no\ grav}}^\star\approx 0.605$ and approaches 1
when $\St\to\infty$. For the smallest value of the Froude number
($\Fr=12$), which is right above the threshold $\Fr^\star\approx 11$,
one clearly observes that small particles collide only if their Stokes
number in a narrow range (here $1.4<\St<3.8$). Again, there are no
backward collisions. They occur only at intermediate values of the
Froude number. As can be seen from Fig.~\ref{fig:efficiency_St} for
the curves associated to $\Fr = 24$, $48$, and $80$, collisions that
occur from the back lead to a non-trivial dependence of $\mathcal{E}$
upon $\St$ with several maxima. At the largest values of the Froude
number, this can lead to collision efficiencies larger than unity. We
will turn back to this behavior in the next two sections.

\begin{figure}[ht]
  \begin{center}
    \includegraphics[width=\columnwidth]{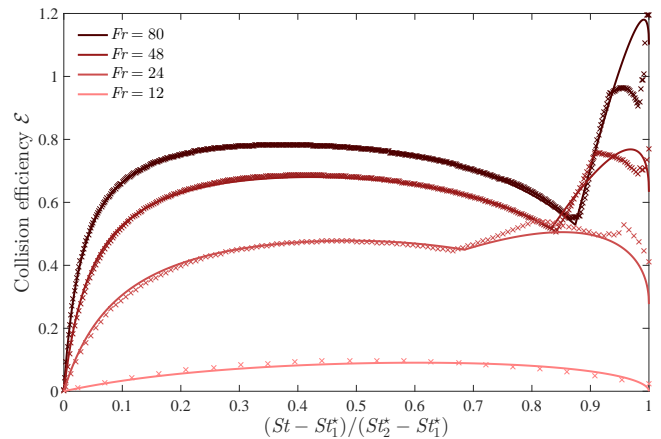}
  \end{center}
  \vspace{-18pt}
  \caption{\label{fig:efficiency_St_fit} Collision efficiencies for
    $\Rey=0$, $\beta=0$, as a function of the reduced Stokes number
    $\sigma = (\St-\St_1^\star) / (\St_2^\star-\St_1^\star)$ for the
    same Froude numbers as Fig.~\ref{fig:efficiency_St}.  Numerical
    data is shown as symbols, while fitting formula are represented as
    solid lines.}
\end{figure}

The efficiency curves as a function of the Stokes number can be
fitted. We use the approximation
\begin{eqnarray}
  &&\mathcal{E}(\St;\Fr) = \mathcal{E}_{\mathrm{front}}(\St;\Fr) +
     \mathcal{E}_{\mathrm{rear}}(\St;\Fr) \ \mbox{with} \nonumber \\
  &&\mathcal{E}_{\mathrm{front}}\approx\frac{
     \alpha_1\,(\St_2^\star-\St_1^\star-3/2)^{2/3}(\St-\St_1^\star)\sqrt{\St_2^\star-\St}}{\St_2^{\star\,4/3}+(\St-\St_1^\star)\,(2\,\St_2^\star-\St_1^\star
     -\St)},\nonumber\\
  &&\mathcal{E}_{\mathrm{rear}}\approx
     \alpha_2\,\St_2^{\star\,0.88}\,\frac{40+15\,\St-13.8\,\St_2^\star}{40+1.2\,\St_2^\star} \label{eq:fit_effic}
\end{eqnarray}
for $\St_1^\star<\St<\St_2^\star$, and where by convention
$\mathcal{E}_{\mathrm{rear}}$ vanishes when the right-hand side is
negative. The constants are adjusted to $\alpha_1 = 0.95$ and
$\alpha_2 = 0.04$ for $\Rey=0$ but might depend on the Reynolds
number. The results of such fits are shown in
Fig.~\ref{fig:efficiency_St_fit}. While the approximation of front
collisions is given by a smooth rational function, the complexity of
backward collisions is reduced to a simple piecewise linear function.

\begin{figure}[h]
  \begin{center}
    \includegraphics[width=\columnwidth]{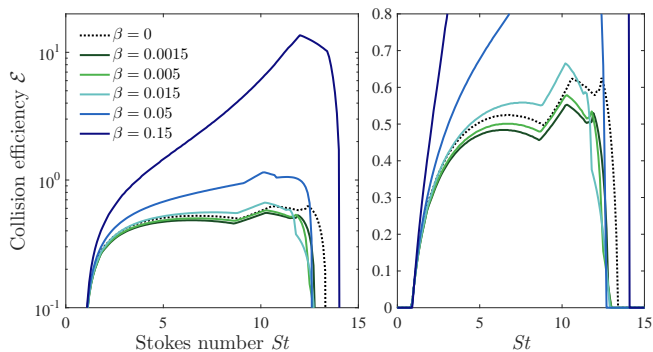}
  \end{center}
  \vspace{-18pt}
  \caption{\label{fig:efficiency_St_beta} Collision efficiencies for
    $\Rey=0$, $\Fr=24$, as a function of the small-particles Stokes
    number and different values of the added-mass parameter $\beta$,
    as labelled. The vertical axis of the left-hand panel is in logarithmic
    units. The right-hand panel show the same data in linear scales. }
\end{figure}

We next turn to investigate the influence on the collision efficiency
of the parameter $\beta$, which controls both the added-mass force and
buoyancy effects.  Figure~\ref{fig:efficiency_St_beta} shows
$\mathcal{E}$ as a function of $\St$ for $\Fr=24$ fixed and various
values of $\beta$ that approximately correspond to mass density ratios
$\rho_{\mathrm{p}}/\rho_{\mathrm{f}} \approx 1000$, $300$, $100$,
$30$, and $10$.  At the smallest values of $\beta$, corresponding for
instance to water droplets in the air, one observes a very tiny
difference with the case $\beta=0$. Collision efficiency are slightly
depleted by less than 10\%.  For $\St$ and $\Fr$ fixed, the effect of
buoyancy is to decrease the settling velocity of small particles. In
principle, this should thus increase their relative velocity with the
collector and lead to higher collision rates. The observed depletion
must hence be due to added-mass effects. This force is proportional to
the fluid acceleration at the particle position. Upstream the
collector, the fluid is decelerated in the direction $z$ and
accelerated in the transverse directions. The particles are thus
pushed aside, decreasing their probability to collide with the sphere.

At larger values of $\beta$, of relevance for instance when
considering slowly sinking organic matter in the ocean, the interplay
between buoyancy and added mass becomes much less trivial. One
observes for the largest values of $\beta$ of
Fig.~\ref{fig:efficiency_St_beta}, that the collision efficiency even
largely exceeds unity, meaning that the flow perturbations are
able to make far-aside particles converge toward the sphere's
trajectory. As we will see in the next section, this somewhat
surprising findings originate from a non-trivial dynamics of the
small particles in the vicinity of the collector.

\section{Shielding and critical Stokes numbers}
\label{sec:shielding}
\begin{figure}[ht]
  \begin{center}
    \includegraphics[scale=.3]{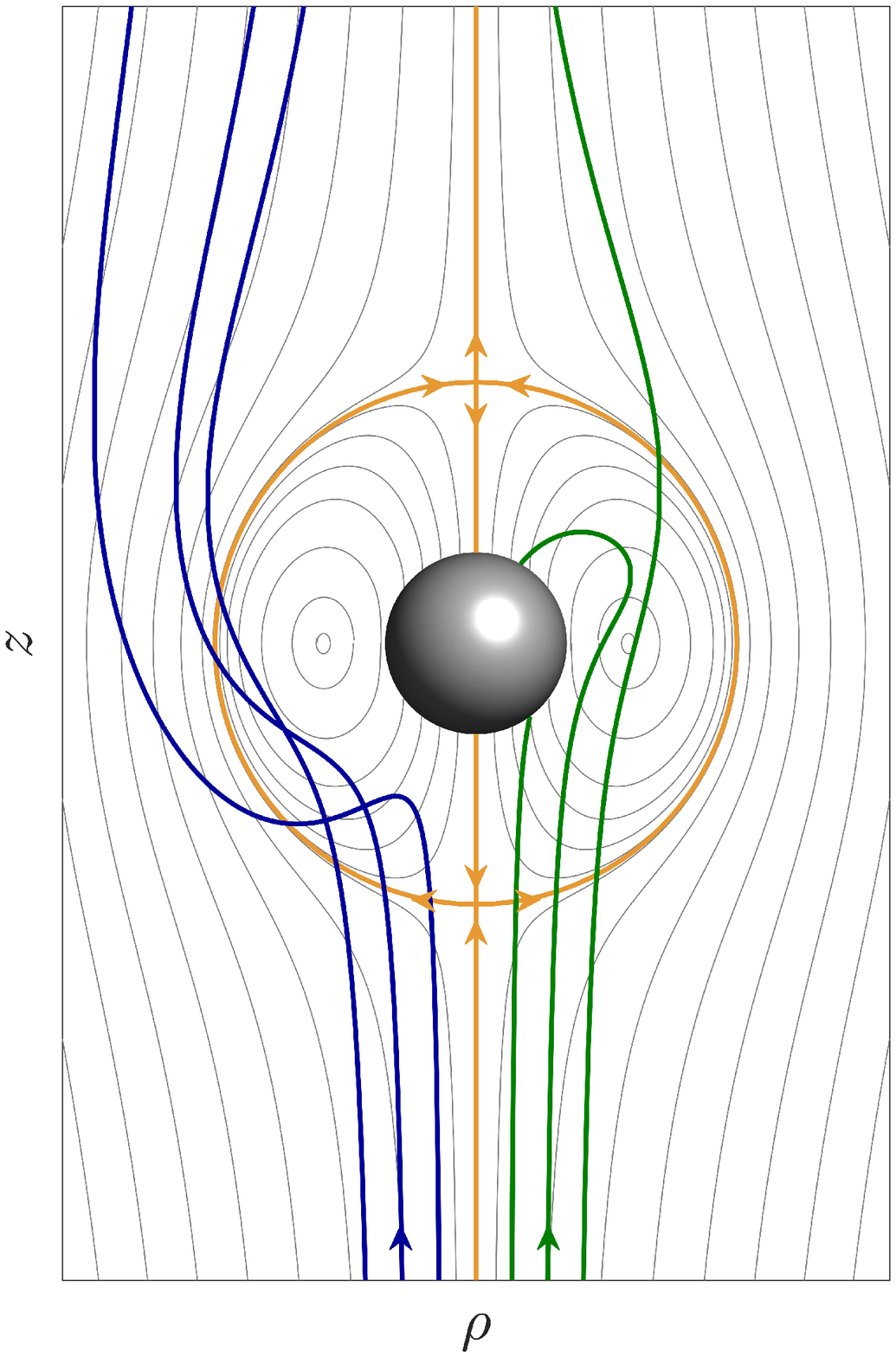}
    \quad
    \includegraphics[scale=.3]{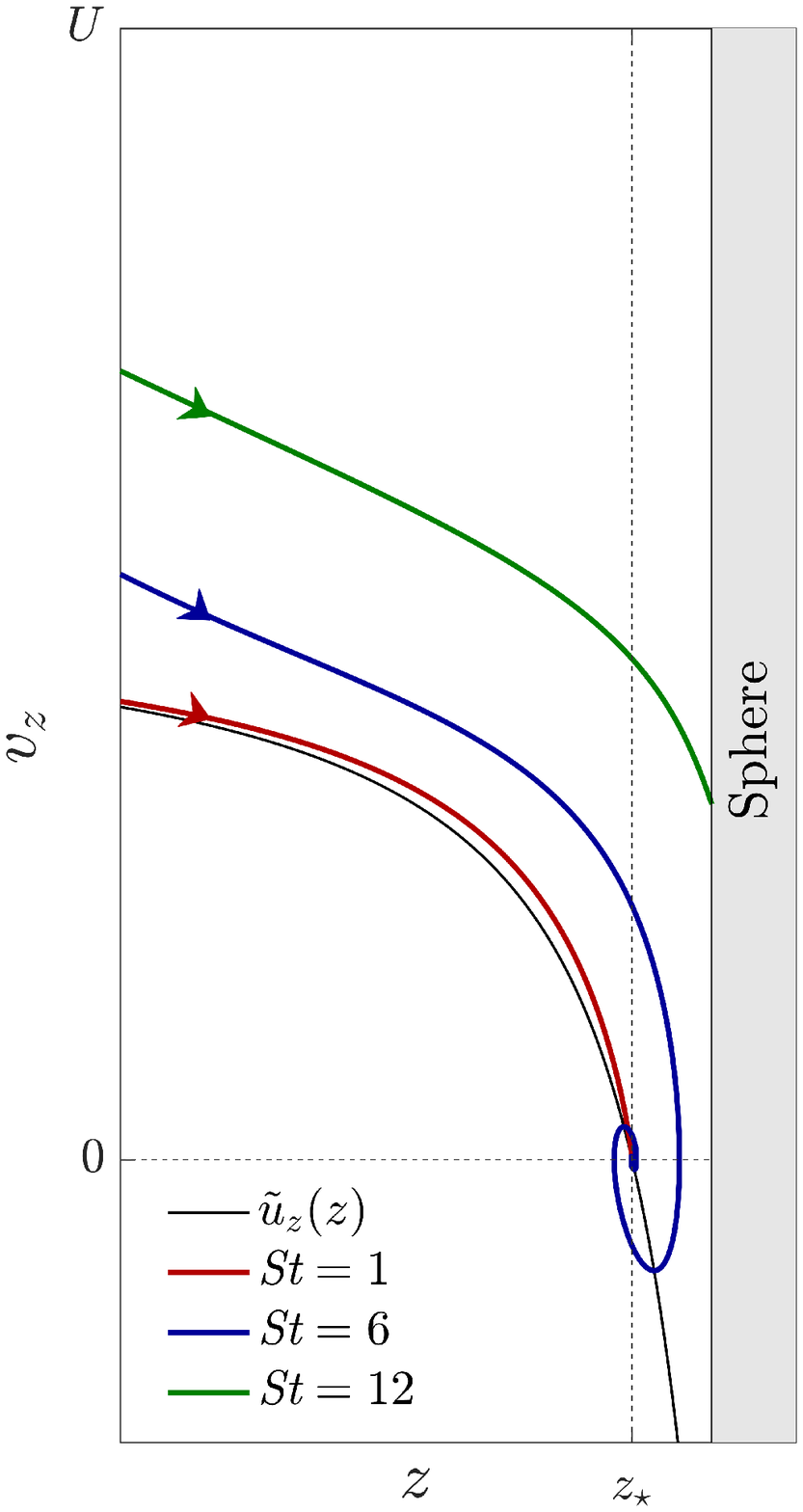}
  \end{center}
  \vspace{-18pt}
  \caption{\label{fig:phase_diagram} Left: Physical-space diagram for
    $\Sg = 1/2$. The gray streamlines show the trajectories of tracers
    of the effective flow $\tilde{\bm u}$. The two saddle fixed points
    are shown in orange, together with their stable and unstable
    manifolds. The bold colored lines represent trajectories
    associated to different $\rho$ at $z=-\infty$ and $\St=6$
    (left-hand, in blue, corresponding to $\Fr=12$) and $\St=12$
    (right-hand, in green, corresponding to $\Fr=24$). Right: Cut of
    the position-velocity phase-space diagram at $\rho=0$ with three
    different trajectories associated to different Stokes numbers and
    satisfying the boundary condition $z = -\infty$ and $v_z = U-\taup\,g$ at
    $t=-\infty$. The thin solid line show the $z$-component of the
    effective fluid velocity profile $\tilde{\bm u}$, which has a
    stagnation point at $z=z_\star$. }
\end{figure}

We have seen in Figs.~\ref{fig:efficiency_St_Fr}
and~\ref{fig:efficiency_St} that, for a given value of the Froude
number, the collision efficiency vanishes outside the interval bounded
by two critical Stokes numbers. We have moreover observed that the
lowest critical Stokes number, $\St_1^\star$ increases when gravity
effects increase (i.e.\ when $\Fr$ decreases), while at the same time,
the upper critical value $\St_2^\star$ decreases.  Let us first focus
on the case $\beta=0$. The effect of gravity is then equivalent to
considering that the small particles are in an effective fluid flow
$\tilde{\bm u} = \bm u + \taup\,\bm g$ given by the sum of the fluid
velocity and their settling speed.  When the gravitational Stokes
number $\Sg$ is less than unity, $\tilde{\bm u}$ has two stagnation
points, at the front and at the rear of the sphere. The left-hand
panel of Fig.~\ref{fig:phase_diagram} shows the streamlines of the
effective flow $\tilde{\bm u}$ represented in the plane of symmetry
$(\rho,z)$, where $\rho^2=x^2+y^2$. The two stagnation points sit on
the $z$-axis of symmetry and are saddle.  The unstable manifolds of
the upstream point are heteroclinic orbits that delimit a
recirculation zone around the particle. Such separatrices act as a
shield around the large sphere. Small particles approaching the
collector are pushed away from it, as for instance illustrated by the
blue trajectories on the left half-plane. For the larger Stokes number
whose trajectories are shown on the right half-plane, the effective
vortices of this recirculation zone are even able to entrain particles
that have sufficiently decelerated to project them toward the back of
the sphere.

When $\beta=0$, the upstream fixed point of the effective velocity
field is located at $(\rho,z) = (0,z_\star)$ where
$u_z(0,z_\star) = \taup\,g$. In Stokes' flow, $z_\star$ is a solution
to
$$
  -\frac{1}{2}\,[\ds/(2z_\star)]^3
  +\frac{3}{2}\,[\ds/(2z_\star)]  +1= \frac{\taup\,g}{U} = \frac{\St}{\Fr}.
\nonumber
$$
This depressed cubic equation has three real roots because
$(1-\St/\Fr)^2-1<0$. Only one of them satisfies $z_\star<-\ds/2$ and
reads $\ds/(2z_\star) = -2\,\sin(\varphi/3)$ with
$\sin \varphi = 1 - \St/\Fr$ and $0\le\varphi\le\pi/2$. One clearly
observes that if $\Fr\to\infty$, then $\varphi\to\pi/2$ and
$\ds/(2z_\star)\to-1$, so that the stagnation point is at the sphere's
surface in the absence of gravity.  If $\Fr\to\St$, then $\varphi\to0$
and $\ds/(2z_\star)\to 0$, so that the stagnation point goes to
$-\infty$. At intermediate values of the Froude number, the shield is
thus at a finite distance from the spherical collector.

This stagnation point, associated with $\vp=0$, defines a fixed point
of the position-velocity particle dynamics. A linear stability
analysis restricted to the plane $\rho=0$, $v_\rho = 0$ gives the two
eigenvalues
$$
  \lambda_{\pm} = \frac{1}{2\,\taup} \left( \pm\sqrt{1 +
      4\,\taup\,\partial_zu_z(z_\star)}-1 \right).
  \nonumber
$$
As $\partial_zu_z(z_\star)<0$, this fixed point is always stable on
the $\rho=0$ manifold. The two eigenvalues are real negative if
$\partial_zu_z(z_\star)>-1/(4\taup)$.  In this case, all trajectories
located upstream on the axis of symmetry converge to the stagnation
point located at $(0,z_\star)$. This is illustrated by the trajectory
associated to $\St=1$ on the right-hand panel of
Fig.~\ref{fig:phase_diagram}. A necessary condition for collisions to
occur is that the particles located on the axis of symmetry reach the
right-hand side of this fixed point (cases $\St=6$ and $\St=12$ in the
right panel of Fig.~\ref{fig:phase_diagram}). This is just
  necessary, but not sufficient, as illustrated by the $\St=6$ case,
  in which the spiralling trajectory is not broad enough to hit the
  sphere. Collisions require that
$$
  \partial_z u_z(z_\star) = \frac{3\,U}{\ds}
  \left[[\ds/(2z_\star)]^4 - [\ds/(2z_\star)]^2\right] < -1/(4\taup),
  \nonumber
$$
so that $\St>1/3$ and
$$
\frac{1}{2} \left(1-\sqrt{1-\frac{1}{3\St}}\right) <
\left[\frac{\ds}{2z_\star}\right]^2 <
\frac{1}{2} \left(1+\sqrt{1-\frac{1}{3\St}}\right).
$$
These two conditions lead to 
\begin{eqnarray}
\Fr &>& \frac{\St}{1-\sin\left[3\,\arcsin
          \sqrt{\frac{1}{8}\left(1-\sqrt{1-1/(3\St)}\right)}\right]},
          \label{eq:cond1}\\
  \Fr &<& \frac{\St}{1-\sin\left[3\,\arcsin
            \sqrt{\frac{1}{8}\left(1+\sqrt{1-1/(3\St)}\right)}\right]}.
            \label{eq:cond2}
\end{eqnarray}
The two corresponding curves are represented in
Fig.~\ref{fig:efficiency_St}. They indeed give bounds on the lower and
upper critical Stokes numbers but can unfortunately not be used as
fits. As they just correspond to a necessary condition, they clearly
stand below the approximation (\ref{eq:fit_stcr}) previously proposed.
The discrepancy is larger for the upper-critical Stokes number for
which the stagnation point is located far upstream the collector. It
is in that case clear that penetrating the shield is markedly not
sufficient to warrant collisions.  Still, the arguments leading to
these two branches explain why there exist two critical Stokes numbers
and show that both stem from similar mechanisms. Below $\St_1^\star$,
as well as above $\St_2^\star$, the unstable manifold of the upstream
saddle stagnation point act as a shield that prevent particles from
penetrating too far in the recirculation zone and thus from hiting the
collector.

\begin{figure}[h]
  \begin{center}
    \includegraphics[scale=.3]{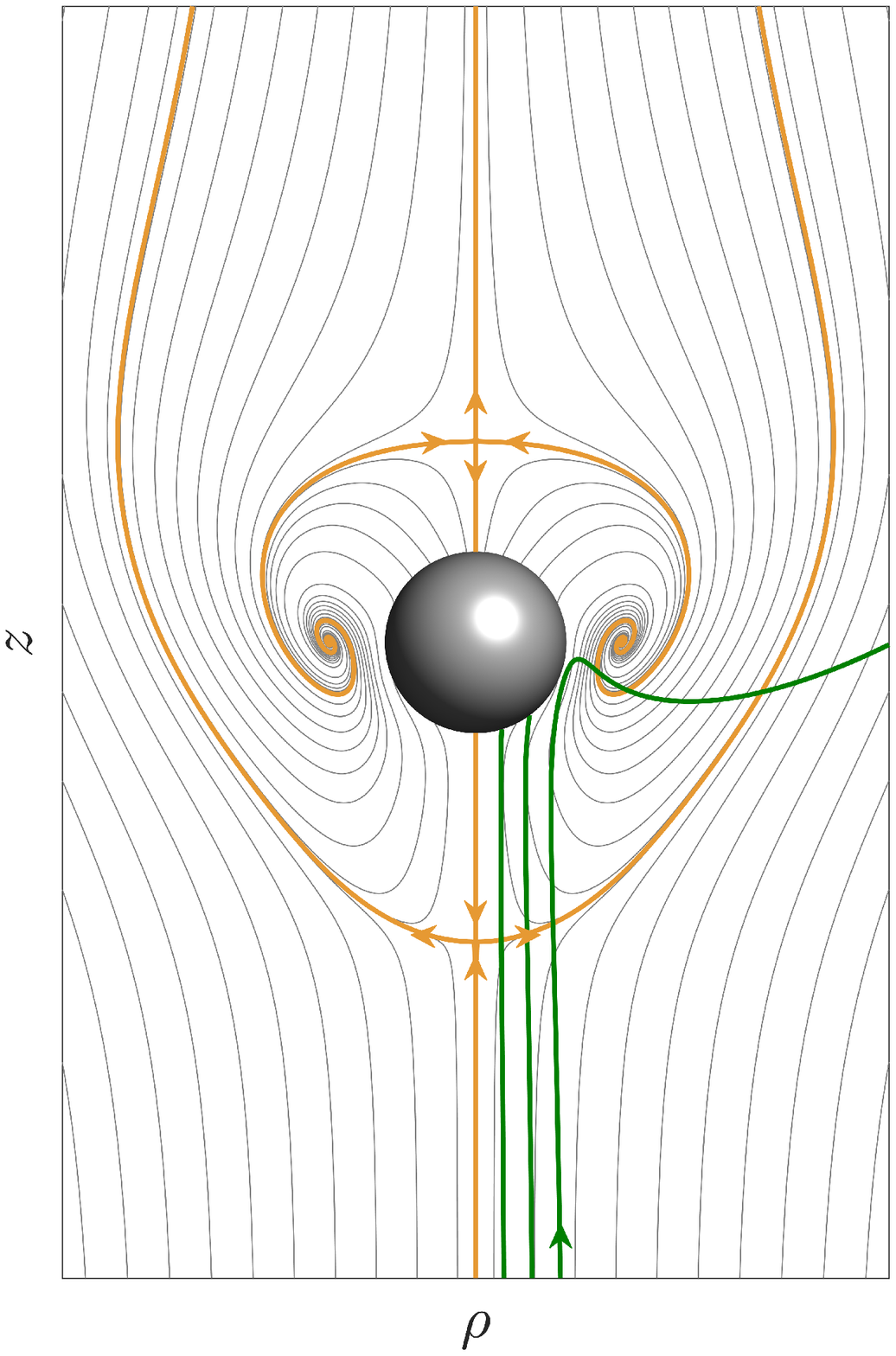}
    \includegraphics[scale=.3]{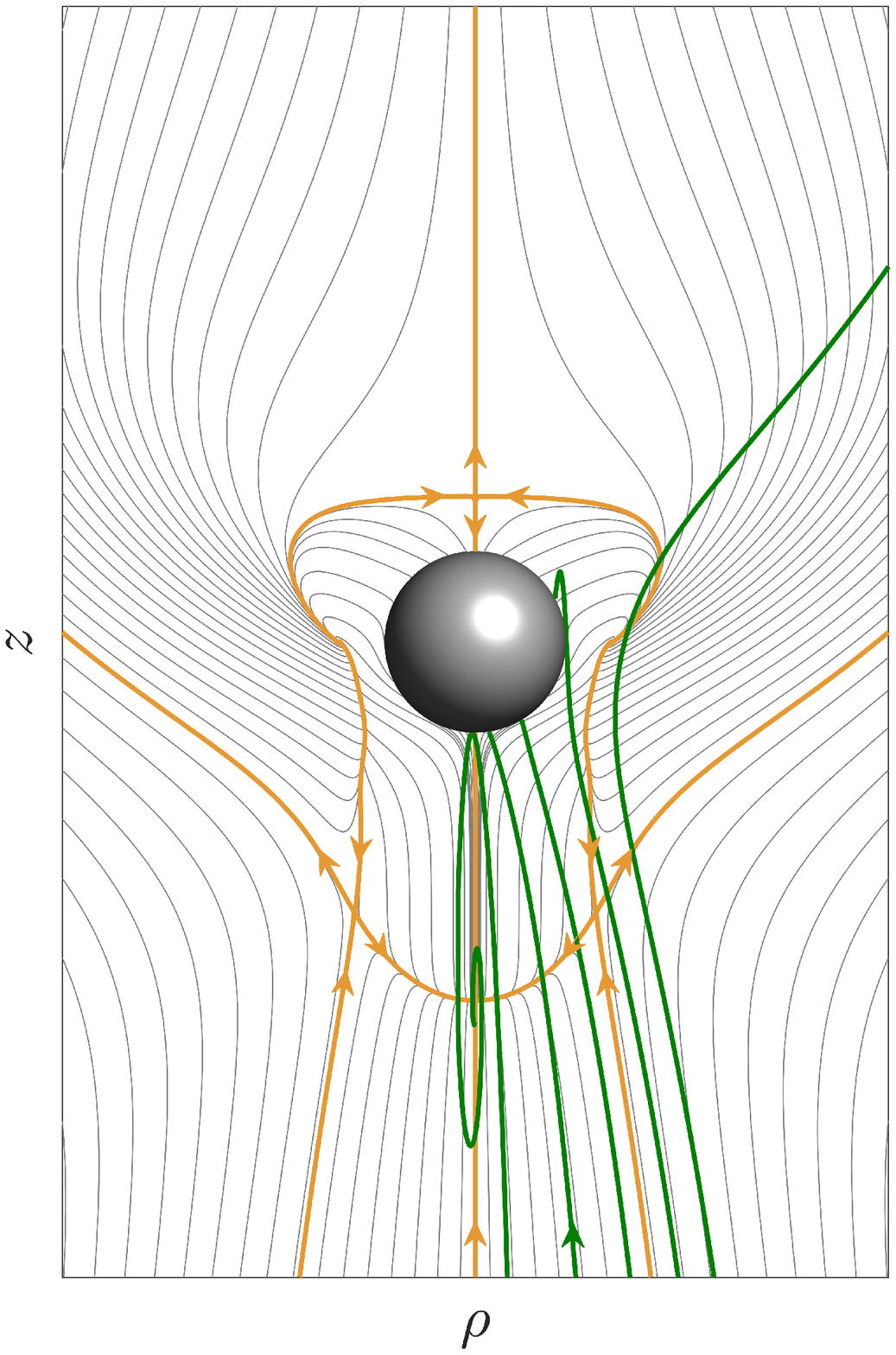}
  \end{center}
  \vspace{-18pt}
  \caption{\label{fig:phase_diagram_beta} Physical-space diagram for
    $\St=12$ and $\Sg = 1/2$ (i.e.\ $\Fr=24$), with $\beta = 0.05$
    (left) and $\beta = 0.15$ (right). The gray streamlines show the
    trajectories of tracers of the effective flow
    $\tilde{\bm u} = \bm u +(1\!-\!\beta)\,\taup\,\bm g +
    \beta\,\taup\,\bm u\cdot\nabla\bm u$. The green curves represent
    trajectories associated to different $\rho$ at $z=-\infty$. }
\end{figure}
It is clear from the considerations drawn above in the case $\beta=0$,
that the near-sphere stagnation points and the associated shield
strongly determines possible collisions of small-size particles with
the collector. We now examine the effects of a finite added-mass
parameter. This time, the particles are as if suspended in an
effective fluid flow, which incorporates the added-mass term, namely
$\tilde{\bm u} = \bm u + (1\!-\!\beta)\,\taup\,\bm g +\beta\,\taup\,
(\mathrm{D}\bm u/\mathrm{D}t)$ and thus adds up a compressible
component. Figure~\ref{fig:phase_diagram_beta} shows the streamlines
of this effective flow for the same parameters as in
Fig.~\ref{fig:phase_diagram}, but with this time $\beta=0.05$ (left
panel) and $\beta=0.15$ (right panel), thus stressing the
modifications of the physical-space diagrams due to a finite added
mass. Differences occur at a qualitative level. First of all, the two
centers that were sitting on the side of the sphere and defining the
effective recirculation flow are now sources.  Moreover, the fore-aft
symmetry is broken.  For the smallest value of the added mass
parameter, the two upstream and downstream saddle fixed points are
still present, but they are no more connected by any heteroclinic
orbit. For the larger value of $\beta$, a bifurcation has occurred and
the upstream saddle point has now become a combination of two saddles
and a sink located on the axis of symmetry $\rho=0$. In both cases,
particle trajectories experience a compression in the transverse
direction and are pushed inward before approaching the sphere,
explaining the efficiencies above unity that are observed in previous
section. Additionally, one observes for the larger value of $\beta$
that the sink present in the effective flow gives rise to a stable
fixed point for the particles dynamics.  Some particles are expected
to get trapped there.

We have seen that the collisions between the small particles
and the collector are largely determined by the near-sphere
dynamics. This process is strongly influenced by the shield-like
structure and the resulting recirculation that appear in the local
topology of the effective flow. Such effects give rise to a variety of
behaviors of the particles trajectories, including quasi-rebounds,
head-on collisions, deflections, trapping and backward collisions.

\section{Caustics and backward collisions}
\label{sec:caustics}

\begin{figure}[b]
  \begin{center}
    \includegraphics[width=\columnwidth]{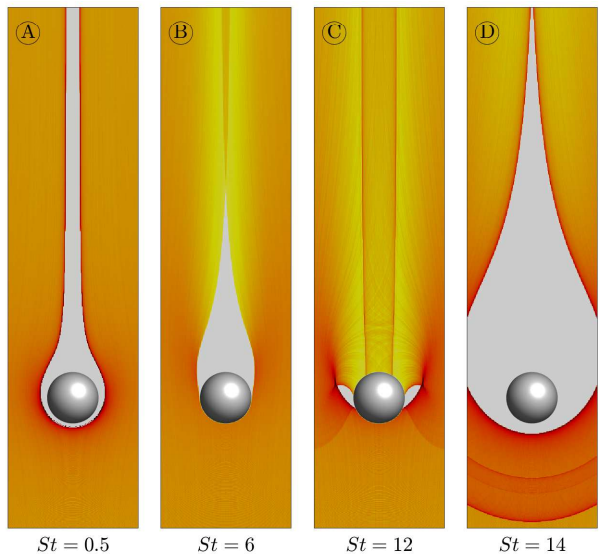}
  \end{center}
  \vspace{-18pt}
  \caption{\label{fig:density_4stokes} Density of particles (colored
    background) for $\Fr = 24$ and four various values of the Stokes
    number, as labeled.}
\end{figure}
Another way to interpret the various behaviours observed in the
previous section consists in drawing, instead of individual
trajectories, the steady-state density of particles with given
characteristics. Figure~\ref{fig:density_4stokes} gives such an
overview for a fixed value of the Froude number, here $\Fr=24$, and
different Stokes numbers representative of the observed regime. In the
case \mycirc{A}, the Stokes number is below the lower critical value
$St_1^\star$. The shield effect prevents small particles from
attaining the collector. They slightly concentrate along an enveloppe
that surrounds the sphere and create a void that extends far away in
the wake. The case \mycirc{B} is representative of what is happening
above the lower critical Stokes number. A fraction of the particles
collides with the sphere, while another is deflected and passes around
the sphere. A void is again created in the wake, but it gets refilled
rather rapidly under the influence of the converging streamlines of
the fluid flow. Still, particles have a rather large inertia that make
them overshoot this tendency, leading to an over-concentration
downstream, along the axis of symmetry. This is a manifestation of the
formation of caustics, and thus of the presence of regions where
particles overlap in space with different velocities.  These
behaviours are still present in the case \mycirc{C} where, in
addition, some particles experience backward collisions. It is clear
from the density profile that such particles are first strongly
decelerated when penetrating the shielded area and are then entrained
by the recirculating effective flow before being pushed back and
collected on the tail of the sphere.  This process creates an
intricate pattern of caustics in the vicinity of the
collector. Finally, case \mycirc{D} is above the upper critical Stokes
number $St_2^\star$. The shield effect again prevents particles from
touching the collector and yields the creation around it of a void
with a tear-drop shape.  This process occurs after several dynamical
rebounds of the particles, suggesting that they oscillate along the
unstable manifold of the upstream stagnation point. This is evidenced
by the presence of several concentric layers in the concentration
profile.

\begin{figure}[h]
  \begin{center}
    \includegraphics[width=\columnwidth]{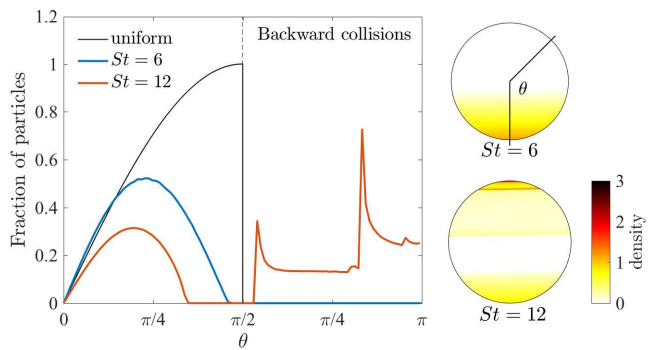}
  \end{center}
  \vspace{-18pt}
  \caption{\label{fig:distrib_pos_coll} Distribution of the angle
    $\theta$ at which the small particles impact the sphere, for cases
    B ($\St=6$) and C ($\St=12$) at $\Fr=24$.}
\end{figure}

To complete the picture, we next turn to quantify further backward
collisions by measuring their spatial spread on the
collector. Figure~\ref{fig:distrib_pos_coll} represents the
distribution of the angle $\theta$ made by the particles position with
$-\bm e_z$ at the instant when they impact the sphere. In case
\mycirc{B}, for which there are no collisions from the back, the
distribution is peaked over small values of $\theta$, and most
collisions occur on the collector's head. In case \mycirc{C}, backward
collisions are dominant. There is no preferential alignment of the
head-on collisions but backward collisions concentrate in several
strips on the downstream hemisphere, with a maximum sitting on the
collector's tail.  Such strong dependences of the angular distribution
of impacts on the small-particle features can have important
consequences on the microphysics of accretion and on the shape
evolution of the collector.

\section{Finite Reynolds numbers}
\label{sec:reynolds}

\begin{figure}[t]
  \begin{center}
    \includegraphics[width=\columnwidth]{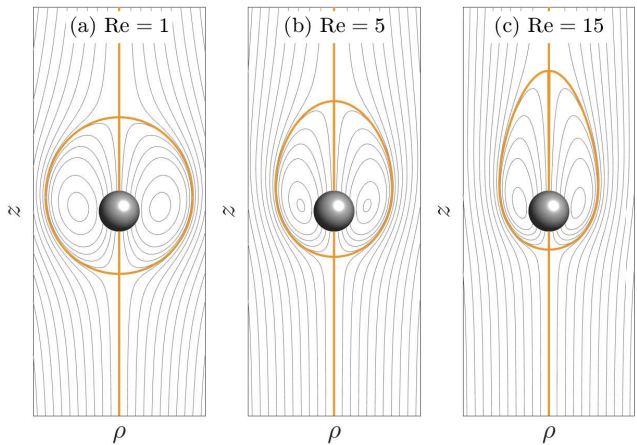}
  \end{center}
  \vspace{-18pt}
  \caption{\label{fig: phase_diagram_diffRe} Physical-space phase
    diagram for $\Sg = 1/2$ (to be compared to the left-hand panel of
    Fig.~\ref{fig:phase_diagram}). The gray streamlines show the
    trajectories of tracers in the effective flow $\tilde{\bm u}$ for
    various Reynolds numbers, as labeled. The bold orange lines are
    the stable and unstable manifolds associated to the two saddle
    stagnation points located upstream and downstream the spherical
    collector.}
\end{figure}

To substantiate the global picture drawn from previous sections, we
here investigate the effect of a finite Reynolds number of the
collector.  At moderate values of $\Rey$ (less than $\approx 15$),
while the fluid velocity field remains axisymmetric, the ``fore-aft''
symmetry observed for Stokes flow $(z,t) \mapsto (-z,-t)$ is
broken. This clearly implies that the two saddle fixed points of the
effective velocity field $\tilde{\bm u}$ introduced in
Sec.~\ref{sec:shielding} are no more symmetric but could also
implicate that they are no more connected by the heteroclinic
trajectories at the origin of the shield effect. The results of
numerical simulations of the Navier--Stokes equation have been used to
reconstruct the physical-space streamlines of the effective velocity
field. The results presented in Fig.~\ref{fig: phase_diagram_diffRe}
in the case $\Sg=1/2$ show that finite, moderate values of $\Rey$ do
not alter the topology of the effective flow. The heteroclinic orbits
are preserved and still bound a well-defined recirculation zone around
the collector. The shape of this region varies when increasing the
Reynolds number and progressively approach an egg-like shape.

\begin{figure}[b]
  \begin{center}
    \includegraphics[width=\columnwidth]{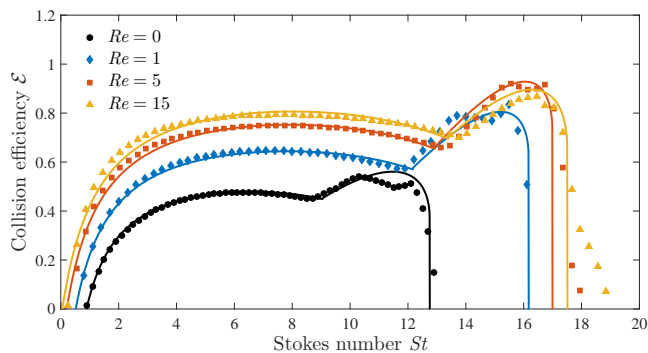}
  \end{center}
  \vspace{-18pt}
  \caption{\label{fig:efficiency_St_diffRey} Collision efficiency at
    $\Fr=24$, as a function of the small particles Stokes number and
    different values of the collector's Reynolds number, as
    labeled. The symbols reports the results of numerical simulation,
    while the solid lines are fits of the form (\ref{eq:fit_effic})
    with parameters given in Tab.~\ref{table}.}
\end{figure}
These qualitative similarities are confirmed when measuring collision
efficiencies for finite values of the collector Reynolds
number. Figure~\ref{fig:efficiency_St_diffRey} represents
$\mathcal{E}$ as a function of $\St$ for the fixed representative
value $\Fr=24$. Two effects are visible when increasing the Reynolds
number. First, the contribution of forward collisions becomes
larger. This is due to sharper variations of the fluid velocity at
$z<0$ and thus the observed shift of the upstream stagnation point
toward the collector. The second observation is that the backward
collisions occurring close to the maximum of $\mathcal{E}$ has a
non-monotonic behavior as a function of the Reynolds number, with a
maximum around $\Rey\approx5$. This comes from a balance between
particles attaining the sphere with a larger speed, and thus less
likely to get captured in the effective recirculation zone, and the
increased extension of this region allowing more particles to being
pushed back toward the collector.  The efficiencies shown in
Fig.~\ref{fig:efficiency_St_diffRey} are fairly well fitted by the
approximation (\ref{eq:fit_effic}) introduced for $\Rey=0$ with
fitting parameters $\alpha_1$ and $\alpha_2$ adjusted as a function of
$\Rey$, as reported in Tab.~\ref{table}.

\begin{table}[h]
  \centering
  \begin{tabular}{c|cccc}
    $\quad\Rey\quad$ & $\quad \alpha_1 \quad$ & $\quad \alpha_2 \quad$
    & $\quad \beta_1 \quad$ & $\quad \beta_2 \quad$\\ \hline
    0 & 0.95 & 0.04 & 1.38 & 0.014\\
    1 & 1.10 & 0.067& 1.22 & 0.0090\\
    5 & 1.25 & 0.078& 1.18 & 0.0059\\
    15 & 1.32 & 0.070 & 1.15 & 0.0042
  \end{tabular}
  \caption{\label{table} Values of the fitting parameters $\alpha_1$
    and $\alpha_2$ used for the approximation (\ref{eq:fit_effic}) of
    the efficiency in Fig.~\ref{fig:efficiency_St_diffRey}, together
    with the parameters $\beta_1$ and $\beta_2$ used for the
    approximation (\ref{eq:fit_stcr}) of the critical lines of
    Fig.~\ref{fig:critic_St_diffRey}, for the various Reynolds numbers
    considered.}
\end{table}

\begin{figure}[h]
  \begin{center}
    \includegraphics[width=\columnwidth]{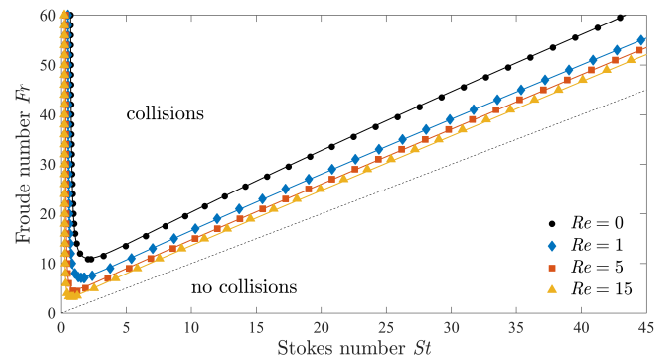}
  \end{center}
  \vspace{-18pt}
  \caption{\label{fig:critic_St_diffRey} Critical lines in the
    parameter space $(\St,\Fr)$ for various values of the Reynolds
    number, as labelled. The symbols are the results of numerical
    simulations, the solid lines are fit of the form
    (\ref{eq:fit_stcr}) with parameters given in Tab.~\ref{table}.}
\end{figure}

To conclude this section, we represent in
Fig.~\ref{fig:critic_St_diffRey} the phase diagram in the $(\St,\Fr)$
parameter space that separates regions with no collisions from those
where $\mathcal{E}>0$.  Numerical results are shown as symbols, while
the solid lines are approximations obtained from the formula
(\ref{eq:fit_stcr}) with parameters $\beta_1$ and $\beta_2$ fitted to
the data. The reported measurements clearly confirm that increasing
the Reynolds number systematically broadens the parameter range over
which collisions occur.

\section{Concluding remarks}
\label{sec:conclusion}

We reported in this paper new findings on the effects of
small-particle gravitational settling on their accretion by a large
spherical collector. Our study shows that such a sedimentation plays a
crucial role in determining collision efficiencies and outcomes of
accretions. The more noticeable conclusions include the presence of a
secondary critical Stokes number above which no collisions occur and
the occurrence of backward collisions where accreted particles are
swept around the spherical collector before falling on its tail.  We
provided a physical interpretation of these phenomena in terms of an
effective velocity field in which the particles are suspended, that is
responsible for the creation of a shield and of an effective
recirculation zone around the collector. 

The effects that we describe should clearly be taken into
consideration in order to improve the parametrization of models used
in atmospheric physics and in astrophysics. Neglecting the
gravitational settling of the small collected particles lead to
drastically misestimating the growth rates of raindrops by collection
or the wet deposition rates of heavy aerosols. Additionally, the
occurence of backward collisions can have an important impact on the
shape of ice crystals during their growth by riming or of planetesimal
in the early Solar system when they accrete dust particles by
filtering.  We provide in this paper fitting formulae both for the
critical Stokes numbers and for the efficiencies that can be of
interest to improve parametrizations.

We finally presented some results on the influence of the
small-particle added-mass forces. Surprisingly, we found that such
effects can be much more significant than those of a finite Reynolds
number of the collector. This is particularly relevant at small values
of the Froude number, or when the mass density ratio between the fluid
and particles cannot be neglected. Added-mass forces could hence be a
key ingredient when estimating the collection rate of sinking organic
matter in the oceans.  The revealed importance of such effects
suggests that added-mass forces are clearly needing more attention and
require further studies.

\
  
We acknowledge C.~Henry and H.~Homann for discussions.  CS thanks the
German Federal Ministry of Education and Research (BMBF) for funding
his project as part of the HD(CP)2 research program under grant
01LK1505B. This paper was initiated during a research stay of CS at
the Observatoire de la C\^ote d'Azur in Nice, which is acknowledged
for its hospitality. CS thanks Germany's National Meteorological
Service, the Deutscher Wetterdienst (DWD), for funding his stay
through the program ``In die Welt f\"ur DWD Know How''.

\bibliography{biblio}

\end{document}